\documentclass[letterpaper,twocolumn,10pt]{article}

\usepackage{usenix}
\usepackage{array}
\usepackage{graphicx}
\usepackage{rotating}
\usepackage{booktabs}
\usepackage{xcolor}
\usepackage{amssymb}
\usepackage{tabularx}
\usepackage{xspace}
\usepackage{soul}
\usepackage{fancyvrb} 
\usepackage{listings}
\usepackage{booktabs}
\usepackage{multirow}
\usepackage{array}

\lstdefinestyle{verbatimstyle}{
  basicstyle=\ttfamily,        
  breaklines=true,             
  breakatwhitespace=false,     
  columns=fullflexible,        
  keepspaces=true,             
  showstringspaces=false,      
  numbers=none,                
  frame=none,                  
  tabsize=2                    
}

\definecolor{terminalbg}{RGB}{245,245,245}   
\definecolor{terminalfg}{RGB}{30,30,30}      
\definecolor{warncolor}{RGB}{200,30,30}      
\definecolor{framecolor}{RGB}{200,200,200}   

\lstdefinestyle{light-terminal}{
    backgroundcolor=\color{terminalbg},
    basicstyle=\ttfamily\small\color{terminalfg},
    breaklines=true,
    breakatwhitespace=false,
    frame=single,
    rulecolor=\color{framecolor},
    showstringspaces=false,
    columns=fullflexible,
    aboveskip=0pt,
    belowskip=0pt,
    lineskip=-1pt,
    emph={CLIENT:, [CLIENT], SERVER:}, emphstyle=\color{warncolor}\bfseries,
    moredelim=**[is][\color{warncolor}\bfseries]{@}{@}
}

\usepackage{epsfig,endnotes}
\usepackage{xurl}
\usepackage{amsmath,amsfonts,chemarrow,balance, graphicx, subcaption, url, multirow, graphics, booktabs, colortbl, xcolor}

\newcommand{\etal}{et~al.\@\xspace}

\usepackage{tikz}

\usepackage{hyperref}
\usepackage[normalem]{ulem}
\usepackage{makecell}
\usepackage{enumitem}
\usepackage{calligra}
\usepackage{aurical}

\newcommand{\BfPara}[1]{{\noindent\bf#1.}\xspace}

\usepackage{pifont}

\newcommand{\scone}{\textbf{\textit{Scenario 1}}\xspace}
\newcommand{\sctwo}{\textbf{\textit{Scenario 2}}\xspace}
\newcommand{\scthree}{\textbf{\textit{Scenario 3}}\xspace}
\newcommand{\scfour}{\textbf{\textit{Scenario 4}}\xspace}
\newcommand{\scfive}{\textbf{\textit{Scenario 5}}\xspace}

\newcommand{\attack}{\textbf{\textit{SpyChain}}\xspace}

\begin{document}
\date{}

\title{SpyChain: Multi-Vector Supply Chain Attacks on Small Satellite Systems}

\author{
{\rm Jack Vanlyssel, Enrique Sobrados, Ramsha Anwar, Gruia-Catalin Roman, Afsah Anwar}\\
Department of Computer Science\\
University of New Mexico\\
Albuquerque, NM, USA\\
\texttt{\{jvanlyssel, esobrados720, ramshaa, gcroman, afsah\}@unm.edu}
}

\maketitle

\begin{abstract}
    
Small satellites are integral to scientific, commercial, and defense missions, but reliance on commercial off-the-shelf (COTS) hardware broadens their attack surface. Although supply chain threats are well studied in other cyber-physical domains, their feasibility and stealth in space systems remain largely unexplored. Prior work has focused on flight software, which benefits from strict security practices and oversight. In contrast, auxiliary COTS components often lack robust assurance yet enjoy comparable access to critical on-board resources, including telemetry, system calls, and the software bus. Despite this privileged access, the insider threat within COTS hardware supply chains has received little attention.

In this work, we present \attack, the first end-to-end design and implementation of an independent and colluding hardware supply chain threats targeting small satellites. Using NASA's satellite simulation (NOS3), we demonstrate that \attack can evade testing, exfiltrate telemetry, disrupt operations, and launch Denial of Service (DoS) attacks through covert channels that bypass ground monitoring. 
Our study traces an escalation from a simple solo component to dynamic, coordinating malware, introducing a taxonomy of stealth across five scenarios. We showcase how implicit trust in auxiliary components enables covert persistence and reveal novel attack vectors, highlighting a new multi-component execution technique which is now incorporated into the SPARTA matrix. Our findings are reinforced by acknowledgement and affirmation from NASA’s NOS3 team. Finally, we implement lightweight onboard defenses, including runtime monitoring, to mitigate threats like \attack.

\end{abstract}
\section{Introduction}

\smallskip
The satellite industry has surged in recent years, reaching an estimated valuation of \$294 billion in 2024~\cite{sia2024ssir}. Much of this growth stems from large-scale constellations like Starlink~\cite{pultarova2025starlink, Tieby2024LEOCyber}, which helped drive a 30\% increase in active satellites over the past year~\cite{clark2025satellite}. Central to this expansion are small satellites—spacecraft with a mass under 180 kilograms~\cite{nasa_smallsats2015}. Their affordability and versatility have reshaped mission planning, enabling deployment at scale for telecommunications, Earth observation, and defense surveillance~\cite{SiddiqueSmallSatRevolution, Kopacz2020,nanoavionics2025smallsats,kogut2024smallsats,Sweeting2018, Song2024}. They are also increasingly used for emerging roles such as satellite internet and autonomous in-orbit experimentation~\cite{SiddiqueEmergingTrends}, underscoring their growing importance across scientific, commercial, and defense domains.

This rapid rise has been driven by design practices that emphasize short development cycles, onboard autonomy, and flexible payloads. Small satellites increasingly adopt multi-mission payload configurations~\cite{burkhard2021evolution}, integrate AI for real-time decision-making~\cite{ortiz2023onboard}, and rely heavily on third-party hardware~\cite{stensberg2024smallsat}. These capabilities are enabled by modular architectures built on core frameworks such as NASA’s Core Flight System (cFS), which provide a stable baseline while allowing peripheral components—sensors, diagnostics, or communication devices—to be added or upgraded as needed. Unlike cFS, which is developed under strict verification and chain-of-custody practices~\cite{NASA-STD-8739.8B,rico2016combineddependabilitysecurityapproach}, external components are not always subject to the same assurance processes.

Similar to terrestrial systems, supply chain compromise is a recognized concern for space mission integrity~\cite{WhiteHouse2025SpaceSystemCybersecurity}. While direct access to satellites limits the number of security studies, researchers have still identified weaknesses in onboard software and hardware, including exploitable firmware flaws~\cite{Willbold2023} and persistent reliance on legacy components integrated for cost and compatibility~\cite{WhiteHouse2025SpaceSystemCybersecurity, falcovacuum}. More concerning is the widespread use of commercial off-the-shelf (COTS) parts, often adopted with little visibility into their provenance or internal logic~\cite{enisa2025spacethreat}. Despite being “auxiliary,” these modules can still subscribe to telemetry, issue commands on the software bus, and manipulate shared memory—granting them deep influence over the flight stack. Prior work has assumed insider modification of core flight software~\cite{donchev2024evaluatingeffectiveransomwareinfection,GuardingGalaxy}, or discussed supply-chain risk in theory~\cite{Peled2023EvaluatingTS}, yet little attention has been given to how malicious functionality could be embedded in COTS modules and persist undetected through integration, launch, and operation. This overlooked vector makes third-party components a particularly plausible pathway for long-term compromise, especially as small satellites proliferate.

To study this overlooked risk, we developed \attack, a supply-chain threat framework deployed in NASA’s Operational Simulator for Small Satellites (NOS3)~\cite{nos3}. Our threat model assumes a supply-chain insider with pre-launch access to third-party hardware components. We argue that such an adversary is realistic considering the prevalence of external vendors and contractors that source individual hardware components, such as sensors, cameras, for satellites~\cite{falcovacuum, Pavur2020SOKBA, Lane2017HighAssurance}. 
Unlike core flight software, auxiliary COTS modules face far fewer assurance requirements, yet enjoy similar access to telemetry, the software bus, and onboard system calls. This asymmetry enables adversaries to establish covert channels, exfiltrate telemetry, or inject arbitrary commands while blending into legitimate message flows. \attack leverages this structural weakness, demonstrating how auxiliary modules can provide a stealthy and persistent foothold that survives integration, launch, and operation.

\attack investigates five attack scenarios. Particularly, we explore single- and multi-component attack scenarios with static and dynamic triggers. For the multi-component attacks, we explored coordination using the software bus and a temporary file to evade telemetry-based ground defenses---attributing stealth to the attack. 
The results validate that hardware supply-chain compromises are feasible with the repurposing of trusted system calls and communication interfaces. 
Additionally, our experiments reveal systemic gaps: small satellites lack software bus authentication, access control, and on-board runtime defense due to hardware constraints. 
To address these weaknesses, we recommend lightweight defenses tailored to cFS and NOS3: runtime behavior monitoring of system calls and message rates, authentication and access control on the software bus, and syscall restriction mechanisms such as seccomp to block covert channels. More broadly we advocate adopting zero-trust module design, supply-chain transparency measures, and operator training to improve resilience against supply chain attacks like \attack.

In summary, we make the following contributions: 

\begin{itemize}[leftmargin=*,itemsep=-1.5pt,topsep=0pt]
    \item \textit{Multi-Level Attacks:} We implement five scenarios that demonstrate different malware architectures and coordination methods, ranging from simple single-component attacks to multi-component attacks using covert file communication. 
    \item \textit{Sensor-Driven Logic:} We show how malware can leverage live satellite telemetry as dynamic triggers for activation or termination. This highlights a realistic pathway for mission-aware, stealthy behavior that goes beyond static, pre-programmed attacks.
    \item \textit{Stealth via Legitimate Interfaces:} We demonstrate that data exfiltration, fault injection, and persistence can be achieved entirely through authorized APIs and system calls, showing how attacks can blend seamlessly with normal operations.
    \item \textit{New Satellite Malware Technique:}
    We mapped all tactics, techniques, and procedures from a successful \attack deployment to the respective standardized strategies listed in the Space Attack Research and Tactic Analysis (SPARTA) matrix~\cite{aerospace2022sparta}. Our investigation found a novel execution technique, using multi-component coordination to achieve a sophisticated stealth attack. After notifying the maintainers, this technique has been added to the SPARTA matrix.
    \item \textit{Actionable Recommendations and NASA Engagement:}  
    Our simulations yielded practical, system-level mitigation strategies tailored to the architecture and constraints of small satellite systems. These findings have been formally acknowledged by the NOS3 development team at NASA, who expressed interest in future collaboration to enhance NOS3 with cybersecurity-focused testbed capabilities.
\end{itemize}
\section{Background and Related Work} \label{sec:back}

This section outlines the technical and contextual foundations necessary to understand \attack. We begin by describing how small satellites are assembled using modular third-party components, and why this integration model introduces risks. We then examine the unique challenges faced by satellite systems. Finally, we analyze prior research on satellite cybersecurity, including simulated attacks and mitigation efforts, and highlight the current knowledge gaps in the literature.

\begin{table*}[]
\centering
\scalebox{0.9}{
\begin{tabular}{llcccc l}
\toprule
 \multirow{2}{*}{\textbf{Prior Work}} & \multirow{2}{*}{\textbf{Targeted Entity}} & \multicolumn{4}{c}{\textbf{Impact}} &\multirow{2}{*}{\textbf{Test Infrastructure}} \\
 \cline{3-6}
 &  & \textbf{DoS} & \textbf{Exfiltration} & \textbf{Persistence} & \textbf{Deception} &\\
\midrule
\hline
Yoon \etal~\cite{Yoon2024} & Inter-satellite Links & \ding{51} & \ding{55} & \ding{55} & \ding{55} & NetSim-3 \\
Willbold \etal~\cite{Willbold2023}* & Firmware & - & - & - & - & QEMU \\
Hansen \etal~\cite{GuardingGalaxy} & Flight Software Stack & \ding{51} & \ding{55} & \ding{55} & \ding{55} &NOS-3 \\
Donchev \etal~\cite{donchev2024evaluatingeffectiveransomwareinfection} & Flight Software Stack & \ding{51} & \ding{55} & \ding{55} & \ding{55} & Raspberry Pi 4B \\ 
\hline
\rowcolor{red!10} \attack & Third-Party Comp & \ding{51} & \ding{51} & \ding{51} & \ding{51}  & NOS-3 \\ 
\hline
\end{tabular}}
 \caption{Comparison of recent papers and attack scenarios with our study. Targeted Entity column details which vulnerability was exploited. Impact column highlights the extent to which an attack can affect the system. Test Infrastructure lists what environment the attacks were simulated in. ${*}$ This study focuses on firmware analysis.}
    \label{tab:literature}
    
\end{table*}

\subsection{The Problem with COTS Components} 
The adoption of COTS components has dramatically lowered the barrier to entry for small satellite development, enabling rapid mission timelines and reduced costs. Yet this economic advantage carries serious risks when third-party hardware is integrated without rigorous verification or oversight. Operators often integrate vendor-supplied hardware that offers little or no visibility into internal logic, privileged system access, or update mechanisms~\cite{gupta2020decentralizedapproachsecurefirmware}. Once in orbit, such components are effectively immutable: unlike terrestrial systems, satellites cannot be physically serviced, are difficult to patch or monitor remotely, and remain exposed to adversaries for the duration of potentially multi-year missions. 

As a result, integrators place disproportionate trust in external suppliers, focusing primarily on ensuring components function as expected rather than verifying their operational logic. Commercial modules typically arrive with an interface specification, binary firmware, and drivers or APIs, but rarely with full source code~\cite{blackbox}. Vendors protect internal logic as proprietary IP, leaving integrators unable to audit firmware beyond interface-level testing. For example, in ESA’s OPS-SAT mission, researchers without source access had to rely on gray-box fuzzing through QEMU emulation~\cite{goehler2022hacking}. 

Without source code, hidden payloads are nearly impossible to detect: static analysis and code review are unavailable, while reverse engineering and fuzzing are costly and beyond the reach of most CubeSat teams~\cite{ruckerl2023distributed}. Resource and schedule constraints further discourage deep inspection, since rapid integration takes priority over adversarial analysis~\cite{nasa2021_smallsat_avionics}.

These blind spots are exploitable. Malware can be fragmented across modules, activated only in coordination, and remain invisible to standard validation. Because incidents are rarely disclosed due to security or reputational concerns, the risks of unvetted third-party components remain systematically underestimated—even when these modules govern critical telemetry, communications, or navigation functions.

\BfPara{Unique Challenges to Satellite Security} 
Currently, a combination of factors make security of satellite systems unique compared to terrestrial cyber-physical systems, e.g. drones. First, satellites support various infrastructures of national importance~\cite{Skelton2023VulnerabilitiesSATCOM}, including 5G/6G connectivity \cite{ESA2023UnlockingConnectivity5G6G}. Satellites hence represent a single point of failure~\cite{falcovacuum} and are increasingly the target of sophisticated cyberattacks, e.g., the 2022 KA-SAT attack that impacted thousands of terminals in Ukraine~\cite{viasat2022ka}. Other attacks have caused irreversible hardware damage and disruption in normal satellite behavior, some of which have been attributed to state actors~\cite{wess2021asat, peters2014noaa, BBCNews}. Despite its significance, space cybersecurity suffers from a regulatory vacuum that has led to a highly fragmented landscape of cybersecurity requirements for space systems~\cite{falcovacuum, Pavur2020SOKBA, WhiteHouse2025SpaceSystemCybersecurity}.

The lack of physical access post-launch~\cite{WhiteHouse2025SpaceSystemCybersecurity} and existence of “blind zones” where the satellite’s orbit takes it outside terrestrial communication links poses another serious challenge due to the impossibility of real-time response to anomalies 
~\cite{CISA2024RecommendationsSpaceSystemCybersecurity, RobertsWeggeman2025OrbitalObservations}. Additionally, inaccessibility post-launch can lead to a satellite being a persistent threat that can endanger other vulnerable space assets. These factors highlight that satellites cannot be secured or managed under the same assumptions as terrestrial systems, requiring distinct approaches to resilient defenses.

\subsection{Prior Research on Satellite Attacks}
The academic and technical literature has explored diverse satellite attack vectors including jamming~\cite{giuliari2021icarus}, spoofing~\cite{oligeri2020gnss,Salkield2023}, firmware analysis~\cite{Willbold2023}, attacks compromising user location~\cite{Jedermann2024RECORD, Liu2025LocationLeakage}, and vulnerabilities in satellite broadband~\cite{pavur2019secrets}, as summarized in Table~\ref{tab:literature}. Much of this work emphasizes radio- and network-level threats—for example, spoofing of downlink channels~\cite{Salkield2023}, denial-of-service attacks on LEO constellations~\cite{giuliari2021icarus}, broadcast storms overwhelming communications~\cite{Yoon2024}, or LTE/5G signaling exploits in direct-to-cell systems~\cite{liu2024dark}—as well as passive exposures such as customer identity leakage~\cite{pavur2019secrets} and exploitation of user terminals~\cite{smailes2023dishingdosdisablesecure}. Complementary defensive research has proposed countermeasures, including location-privacy protections for satellite internet users~\cite{Koisser2024DontShoot} and risk-scoring frameworks for space systems~\cite{AnjumFarheen2025SoKSpaceCybersecurity}. Relatively little attention has been devoted to vulnerabilities in firmware and onboard software.

Donchev \etal~\cite{donchev2024evaluatingeffectiveransomwareinfection} and Hansen \etal~\cite{GuardingGalaxy} both created proof-of-concept ransomware for satellites. These works highlight the potential impact of malicious onboard code, but they assume highly privileged access. Specifically, both models require insider capability to edit and recompile the flight software itself—either via a NASA insider with rights to modify the publicly maintained cFS baseline, or an insider embedded within a mission’s flight-software or launch team with commit rights to the flight build.

This path is implausible in practice. The cFS baseline is maintained under strict public configuration control, meaning any malicious edits would be visible. Similarly, compromising a mission's flight build would require inserting a trusted insider into a launch or operations team, roles with multiple layers of review and oversight. While not impossible, this insider vector is far less practical than often portrayed, especially when compared to the more plausible supply chain path, where third-party components are purchased and then installed on top of the core flight software.

Supply chain compromise is a recurring theme in satellite cybersecurity~\cite{Peled2023EvaluatingTS,kapalidis2019cyber,salim2024cybersecurity,botezatuspace,falcovacuum,Manulis2021}, but most discussions remain theoretical or limited to broad risk assessments. While some propose architectures for risk reduction~\cite{sawik2023space,yadav2024orbital}, they do not examine malware operating within actual flight software stacks. Our work addresses this gap by simulating coordinated supply-chain malware in NASA’s NOS3 environment, where vendor-supplied third-party components integrate as plug-and-play additions. We show how such components can collude to establish covert channels and exfiltrate telemetry while maintaining legitimate-appearing behavior, enabling systematic study of persistence, stealth, and coordination in a realistic system context and motivating countermeasures grounded in observed behavior rather than conceptual models.
\section{Threat Model} \label{sec:model}

This section defines the important facets that make \attack realizable. We begin by describing our adversary model, the satellite attack surface, and the assumptions and their validity. 

\subsection{Adversary Model}

We consider a \textit{Supply Chain Insider} with pre-launch access to hardware components via vendors. The adversary is assumed to be familiar with the target satellite’s flight software and operating system, which enables them to build functional malicious code. Specifically, the adversary has knowledge of how telemetry is structured and routed, and how communication channels integrate with the flight software. NASA's flight software (cFS) is the most commonly used flight software and is open-source, reducing the knowledge barrier~\cite{werner2025cfsupdate}. Knowledge of the onboard operating system (e.g., Linux, RTOS) is also needed to access standard system interfaces such as interprocess communication and network calls. These interfaces allow the creation of covert communication channels. Some of this knowledge gap is naturally bridged during component development, since hardware vendors must design modules for a specific operating system and flight software environment, giving them detailed insight into how those systems function.
Although the barrier for manufacturing a hardware component is very high, the presence of contractors and the high success rate have encouraged multiple similar threat actors including nation states~\cite{SASC2012Counterfeit,NYTimes2024PagerExplosions}. Acknowledging these problems, supply chain-induced threats in general have been identified as a priority by the FCC~\cite{fccSupplyChain}. 

\begin{figure}[t]
    \centering
    \includegraphics[width=0.75\linewidth]{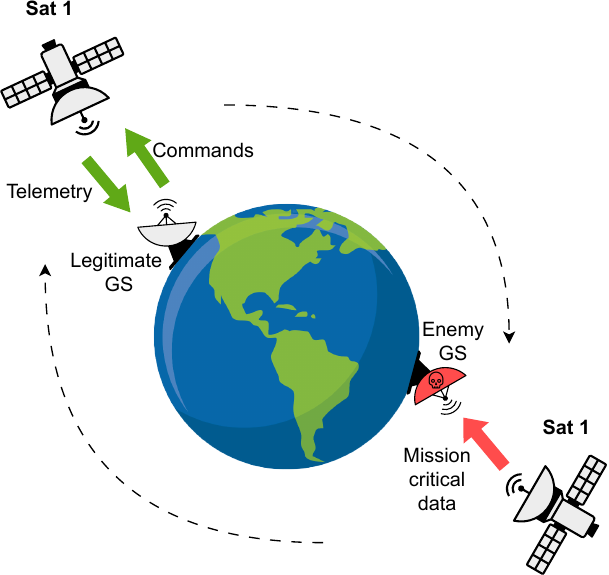}
    \caption{The same satellite passing over both a legitimate ground station and a malicious ground station.}
    \label{fig:sat_architecture}
\end{figure}

In addition, we assume the adversary has access to a ground station positioned within the satellite’s orbital path, such that the spacecraft will periodically pass overhead. This capability enables the attacker to receive exfiltrated telemetry once the malware activates. The availability of software-defined radios and community projects such as SatNOGS~\cite{pavur2019secrets} further reduces the barrier to intercepting and processing satellite downlinks. Figure~\ref{fig:sat_architecture} illustrates this threat model, contrasting the legitimate ground station with the adversary’s station that covertly receives mission-critical data.

\subsection{Attack Surface}

The attack surface of small satellites stems from their structural design: modular applications share broad, implicitly trusted system resources without fine-grained isolation. This creates opportunities for a malicious component to subvert communication, storage, and transmission channels that are assumed to be safe.

\begin{itemize}[leftmargin=*,itemsep=-3pt,topsep=0pt]
    \item \textit{Communication Channels:} The software bus facilitates internal satellite communication and allows any onboard application to send or receive telemetry and commands without authentication. The radio facilitates external communication and allows unauthenticated connections, enabling data to be downlinked outside monitored telemetry paths. 
    \item \textit{System Interfaces:} Flight software environments leave common abstractions such as file systems and OS interfaces exposed. A malicious component can exploit these to establish covert storage or communication channels.
    \item \textit{Logging and Forensic Gaps:} Small satellites lack the logging and runtime visibility common in ground systems~\cite{ RobertsWeggeman2025OrbitalObservations}. System calls, file operations, and network use are not logged due to hardware constraints, and any logs that are generated are often volatile~\cite{YuanTelemetry2023}. This absence of monitoring prevents operators from detecting anomalies or conducting effective forensic analysis.
    \item \textit{Operational Constraints:} Satellites deliberately sacrifice security controls for power, mass, and bandwidth efficiency. Memory and CPU limitations preclude real-time anomaly detection, syscall filtering, or sandboxing, while downlink constraints prevent operators from streaming raw forensic data to the ground for analysis.
    \item \textit{Lack of Security Mechanisms:}  Unlike terrestrial systems, small satellites have "almost nothing" in terms of dedicated intrusion detection, access control, or runtime validation tools~\cite{burgess2023satellites}. In our correspondence, NASA has acknowledged that, currently, cFS missions lack such defenses, validating that resilience is mainly dependent on trust in component integrity and pre-launch analysis.
\end{itemize}

\subsection{Stakeholders}
An attack like \attack can inflict massive damage to a wide range of stakeholders. For vendors, it creates liability and reputational risk as their satellite components may become vectors for compromise. For flight software developers, \attack exposes how malicious logic can blend into legitimate message flows, undermining confidence in assurance processes and the trustworthiness of the software framework itself. Satellite Developers face the risk of integrating compromised components into the flight system. A missed backdoor or hidden payload can undermine confidence in their integration process and damage their reputation as a trusted builder. Mission Operators bear the operational consequences. They are responsible for flying and maintaining the satellite day-to-day, meaning any covert compromise directly threatens mission success, data integrity, and operator trust. End Users (including researchers, customers, or defense analysts) ultimately feel the downstream effects. They may face degraded data integrity, mission disruption, or leakage of sensitive payload results, often without ever realizing the satellite was compromised in the first place.

\begin{figure*}[t]
  \centering
  \begin{subfigure}[b]{0.275\linewidth}
    \includegraphics[width=\linewidth]{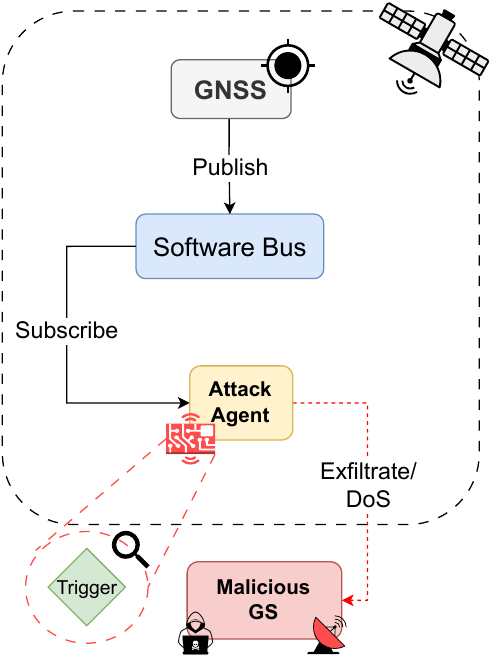}
    \caption{Single malicious Component.}
    \label{fig:scenario_single_component}
  \end{subfigure}
  \hfill
  \begin{subfigure}[b]{0.35\linewidth}
    \includegraphics[width=\linewidth]{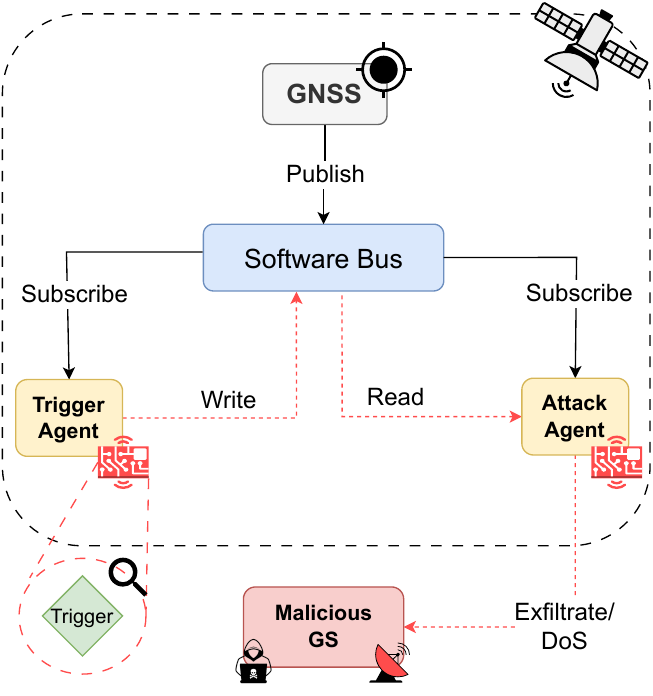}
    \caption{Multiple malicious components.}
    \label{fig:scenario_multi_component}
  \end{subfigure}
  \hfill
  \begin{subfigure}[b]{0.35\linewidth}
    \includegraphics[width=\linewidth]{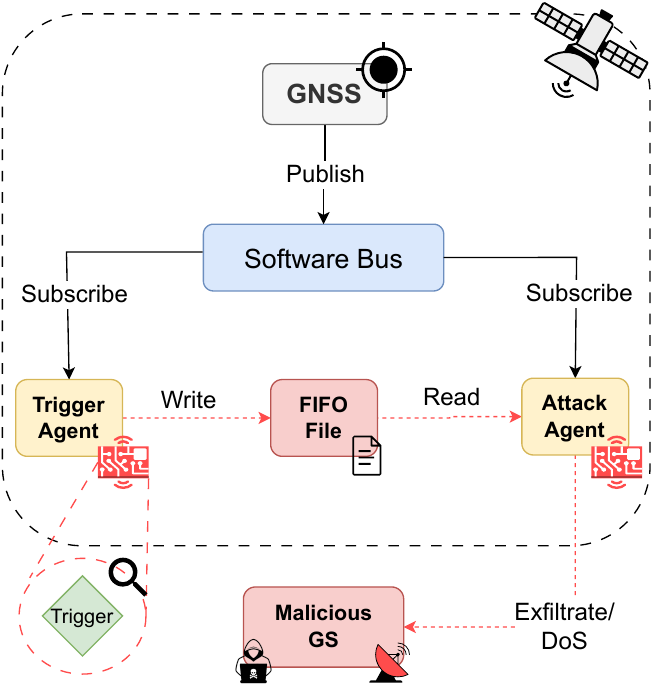}
    \caption{Multiple malicious components with FIFO.}
    \label{fig:scenario_multi_file}
  \end{subfigure}
  \caption{Attack architectures showing solo and colluding components. The attack pipeline is depicted in red. The GNSS sensor is a component which publishes location measurements which may be used as a trigger. The trigger can also be time based. When triggered, the attack agent can either exfiltrate data or perform a DoS attack. \ref{fig:scenario_single_component} shows the attack pipeline using one component (\scone and \sctwo); \ref{fig:scenario_multi_component} for multiple components (\scthree and \scfour); and \ref{fig:scenario_multi_file} for multiple components using a file as the communication channel (\scfive).}

  \label{fig:attack_scenarios}
\end{figure*}

\section{\attack Methodology} \label{sec:method}

In this section, we present \attack. We detail the attack strategies, scenarios, and timeline from activation to execution. We then describe the concrete implementation of the malware agents, showing how they were constructed, coordinated, and integrated into the simulated satellite environment.

\subsection{Attack Objectives}
The \attack attack framework was designed to demonstrate three core adversarial objectives that a supply chain attacker could pursue in a small satellite environment: data exfiltration, denial of service, and persistence. 

\begin{itemize}[leftmargin=*,topsep=0pt,itemsep=0pt]
    \item \textit{Denial of Service:} A sustained DoS can simulate plausible system faults, suppress incident response, or degrade operations during critical mission phases. For an attacker, the ability to silence or disable a satellite at will represents one of the most dangerous threats to mission success. This severity explains why so many prior studies focus on DoS-style scenarios, often as the baseline case for demonstrating space system vulnerability.
    \item \textit{Data Exfiltration:}  Gaining continuous access to mission telemetry allows adversaries to map spacecraft health, payload operations, and orbital state in unprecedented detail. This information has dual value: it can be weaponized for espionage or reconnaissance, and it enables precision follow-on attacks by revealing exploitable system weaknesses. Unlike transient DoS effects, exfiltration builds long-term intelligence advantages.
    \item\textit{Persistence:} For an attacker, maintaining a foothold across the entire mission lifecycle is as valuable as any single disruption. Malware that survives integration, launch, and on-orbit operations—by hiding in trusted components and activating only under mission-relevant triggers—ensures uninterrupted access. This persistence eliminates the need for reinfection and ensures a lasting influence channel for the adversary.
    \item \textit{Disruption:} The ability to subtly manipulate telemetry or inject deceptive commands offers adversaries the opportunity to erode operator trust in the spacecraft. False data streams can mask exfiltration, trigger wasteful responses, or degrade mission objectives without a single destructive act. Disruption complements persistence and exfiltration, giving attackers the power not just to observe but to reshape mission outcomes.
\end{itemize}

\subsection{Attack Architecture}

Across our simulated attack scenarios, several common design principles emerged that enabled stealth, coordination, and data exfiltration within the constraints of our simulated satellite. This section describes the shared architectural elements and malicious techniques that were foundational for all scenarios.

\textbf{Legitimate Appearance:} The malicious components were implemented as standard cFS applications, compliant with typical development practices and integrated into the onboard software stack. All scenarios passed integration and testing, and used standard communication channels for their intended task to avoid scrutiny from operators.

\textbf{Use of Native OS System Calls:} To maintain compatibility and avoid detection, the malware leveraged familiar system calls available in Linux-based embedded environments. The malicious components used these system calls to create communication channels and exfiltrate data using the radio. Such calls remained unseen to operators.

\textbf{Modular Design:} Each scenario was realized through one or more applications that could be independently integrated into the flight stack. This modular design mirrors real satellite software practices, where components from multiple vendors are combined. It enabled exploration of single versus multi-component attacks, coordination via files or the software bus, and both static and dynamic activation strategies. Such flexibility illustrates how an adversary could adapt an attack to a specific mission or distribute malicious logic across modules to increase obfuscation.

\textbf{Reuse of Legitimate Subscriptions:}
In all attacks, the malicious applications subscribed to legitimate telemetry streams—such as GNSS navigation or electrical power system data—using the same mechanisms as trusted components. Because cFS provides no authentication or access control on subscriptions, these actions were indistinguishable from normal operation and invisible to system integrators. As a result, operators remained unaware that additional components were siphoning mission-critical telemetry, which was then exfiltrated through covert channels.

\textbf{Trigger Based Activation:}
All malicious components were designed to perform only expected functions during early mission phases, including integration, testing, and launch. Each scenario employs an activation mechanism ranging from time-based delays to sensor-driven conditions, ensuring that exfiltration only begins once a predefined trigger occurs. This strategic delay enhances stealth by preventing premature detection.

\subsection{Attack Scenarios}

We implemented five distinct attack scenarios, each escalating in architectural complexity. These variations demonstrate how even minimal capabilities can lead to successful data exfiltration when implicit trust and architectural assumptions are exploited. Figure~\ref{fig:attack_scenarios} depicts the \attack\ pipeline under these scenarios.

\paragraph{Scenario 1 --- Single Component, Static Time Trigger}
\begin{itemize}[leftmargin=*,itemsep=-3pt,topsep=0pt]
    \item \textit{Architecture:} One malicious cFS application integrated into the satellite software stack.
    \item \textit{Trigger Logic:} A fixed countdown timer starting at system boot.
    \item \textit{Coordination Method:} None; the application operates independently.
    \item \textit{Attack Flow:} After initialization, the application waits for the countdown to expire, then begins sending mission telemetry directly to the malicious ground station using the onboard radio.
\end{itemize}

\begin{figure*}[t]
    \centering
    \includegraphics[width=0.9\textwidth]{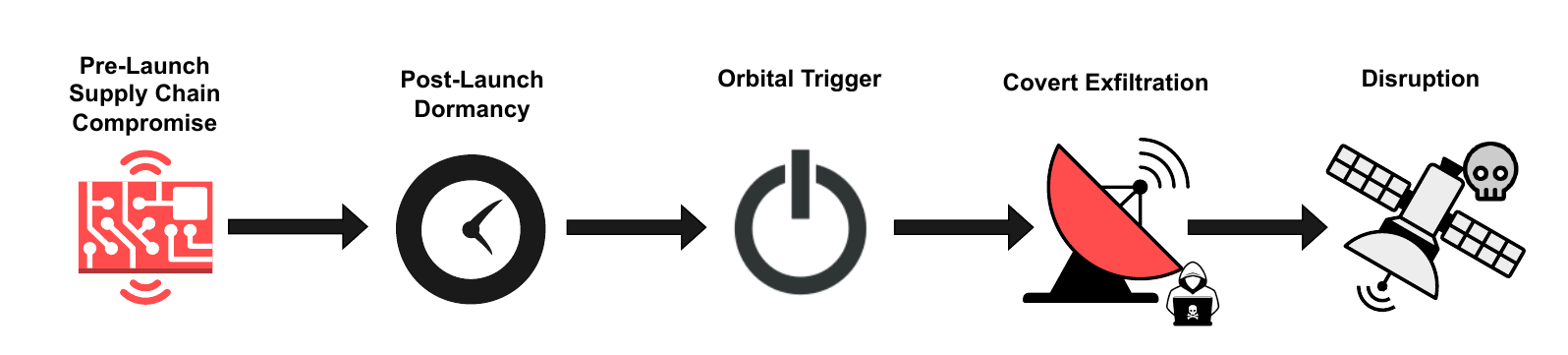}
    \caption{Attack timeline overview for \scfive: (1) malicious software is embedded within the components, (2) after launch the trigger agent waits for a trigger condition, (3) the trigger agent writes commands to the FIFO file and the attack agent read them, (4) attack agent exfiltrates the mission data to malicious GS, (5) malware ceases exfiltration and performs other attacks.
    }
    \label{fig:attack_timeline}
\end{figure*}

\paragraph{Scenario 2 --- Single Component, Dynamic Sensor Trigger}
\begin{itemize}[leftmargin=*,itemsep=-3pt,topsep=0pt]
    \item \textit{Architecture:} One malicious cFS application with dynamic activation.
    \item \textit{Trigger Logic:} Subscription to GNSS telemetry; activation occurs upon detection of orbital conditions.
    \item \textit{Coordination Method:} None; the application operates independently.
    \item \textit{Attack Flow:} After initialization, the application monitors the GNSS telemetry stream. Once GNSS activity indicates orbital insertion, the application begins transmitting mission telemetry using the onboard radio.
\end{itemize}

\paragraph{Scenario 3 --- Multi Component, Static Trigger, Software Bus Coordination}
\begin{itemize}[leftmargin=*,itemsep=-3pt,topsep=0pt]
    \item \textit{Architecture:} Two malicious cFS applications---a trigger agent and an attack agent---integrated separately into the flight software.
    \item \textit{Trigger Logic:} A fixed countdown timer in the trigger agent.
    \item \textit{Coordination Method:} Messaging over the cFS software bus.
    \item \textit{Attack Flow:} After initialization, the trigger agent waits for the countdown to expire, then publishes a predefined coordination message on the software bus. The attack agent, subscribed to this message, begins transmitting mission telemetry to the malicious ground station.
\end{itemize}

\paragraph{Scenario 4 --- Multi Component, Dynamic Trigger, Software Bus Coordination}
\begin{itemize}[leftmargin=*,itemsep=-3pt,topsep=0pt]
    \item \textit{Architecture:} Two malicious cFS applications---a trigger agent and an attack agent---integrated separately into the flight software with dynamic activation.
    \item \textit{Trigger Logic:} Subscription to GNSS telemetry in the trigger agent; activation occurs upon detection of orbital conditions.
    \item \textit{Coordination Method:} Messaging over the cFS software bus.
    \item \textit{Attack Flow:} The trigger agent monitors GNSS telemetry until orbital insertion is detected, then sends a coordination message over the software bus. The attack agent, upon receiving the message, starts transmitting mission telemetry to the malicious ground station.
\end{itemize}

\paragraph{Scenario 5 --- Multi Component, Dynamic Trigger, File-Based Coordination}
\begin{itemize}[leftmargin=*,itemsep=-3pt,topsep=0pt]
    \item \textit{Architecture:} Two malicious cFS applications---a trigger agent and an attack agent---integrated separately into the flight software with dynamic activation.
    \item \textit{Trigger Logic:} Subscription to GNSS telemetry in the trigger agent; activation occurs upon detection of orbital conditions.
    \item \textit{Coordination Method:} Writes and reads to a temporary file in the satellite file system.
    \item \textit{Attack Flow:} The trigger agent monitors GNSS telemetry until orbital insertion is detected, then writes an activation command to the FIFO file. The attack agent polls the file at regular intervals and, upon reading the activation command, begins transmitting mission telemetry to the malicious ground station using the onboard radio.
\end{itemize}

\subsection{Attack Timeline}

Although our simulation explored five distinct attack variants with increasing complexity, this section focuses on \scfive. We selected this scenario for detailed analysis because it best illustrates how supply chain malware can operate undetected using trusted interfaces and covert communication channels. Figure~\ref{fig:attack_timeline} illustrates this timeline and is discussed in detail below.

\begin{enumerate}[leftmargin=*,itemsep=-3pt,topsep=0pt]
    \item \textbf{Supply Chain Compromise:} The attack begins during the satellite's integration phase when multiple malware-infected components are installed into the satellite's software stack. These modules pass through testing due to their legitimate appearance and behavior. 

    \item \textbf{Dormant Operation:} After launch, the components perform expected duties while malicious logic lies idle. The trigger agent polls the GNSS telemetry stream for orbital indicators, while the attack agent continuously polls the FIFO file. No suspicious activity is observable at this stage.
    
    \item \textbf{Trigger Activation:} When the trigger agent detects that the satellite has entered orbit (from GNSS activity), it writes a signal to the shared file. This write operation is indistinguishable from file access patterns in many embedded systems.
    
    \item \textbf{Data Exfiltration:} The attack agent component continues to poll the file, and once it reads the signal, the attack agent begins to transmit telemetry. The exfiltrated data is routed through the radio, so it occurs without ground control's knowledge.
    
    \item \textbf{Termination or Persistence:} Depending on the attacker’s goals, the \attack can crash the flight software with a segmentation fault or flood the software bus, enabling denial-of-service. It can also remain active indefinitely by using triggers to persistently exfiltrate telemetry or generate confusion by publishing arbitrary messages and commands on the bus. In practice, these behaviors can be combined or tailored, allowing the attacker to choose between disruption, stealthy persistence, or operator deception.
\end{enumerate}

This timeline illustrates how a multi-component malware system can blend into standard spacecraft operations, activate only under mission-relevant conditions, and exfiltrate sensitive data using trusted mechanisms without triggering alarms in standard ground telemetry systems.

\begin{figure}[t]
  \centering
  \begin{subfigure}[b]{0.45\linewidth}
    \includegraphics[width=\linewidth]{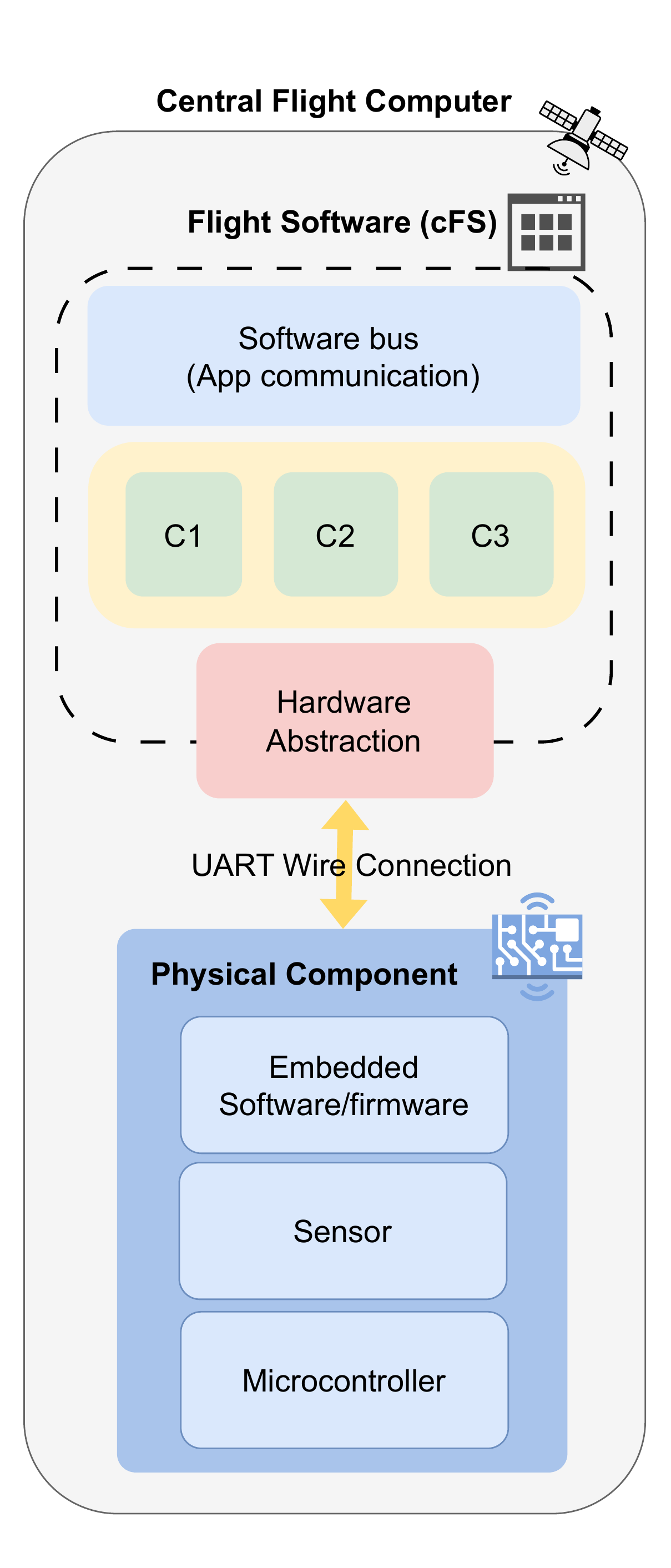}
    \caption{Physical Satellite}
  \end{subfigure}
  \hfill
  \begin{subfigure}[b]{0.45\linewidth}
    \includegraphics[width=\linewidth]{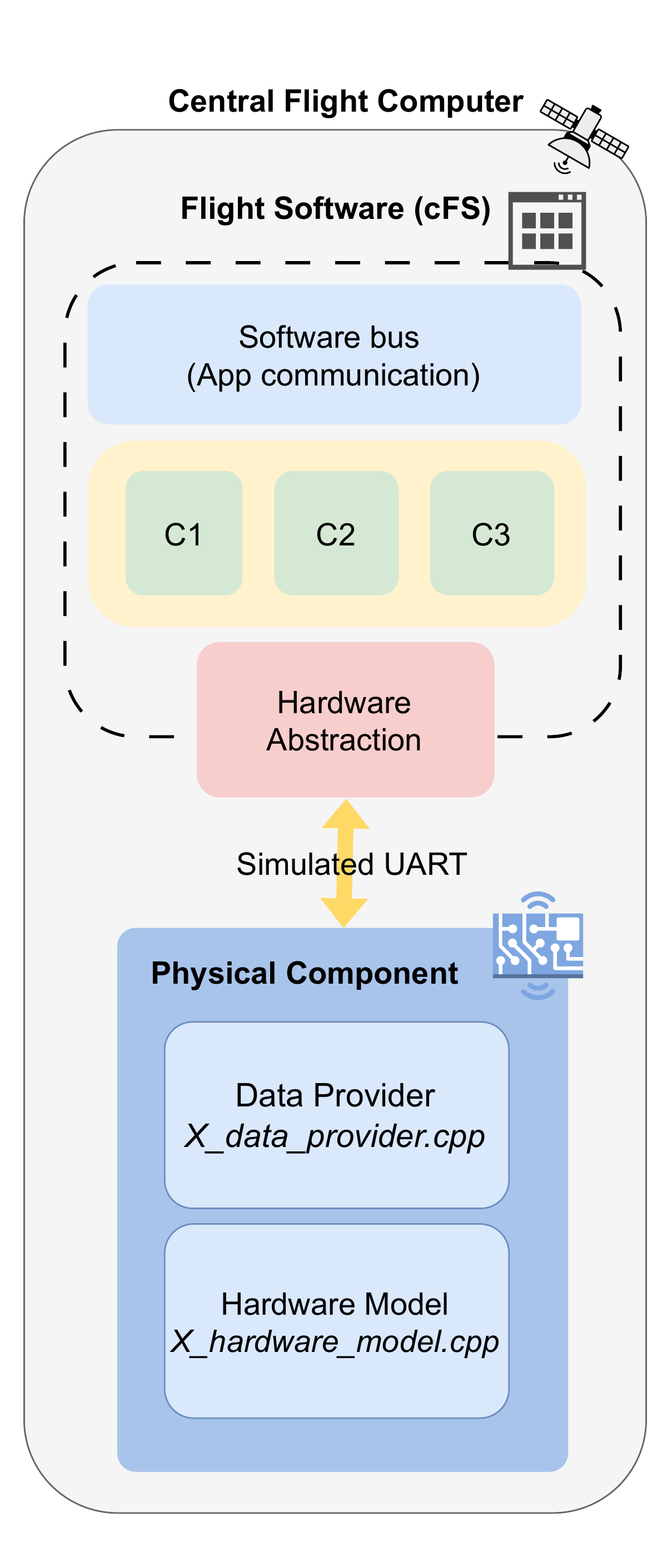}
    \caption{NOS3 Simulated}
  \end{subfigure}
  \caption{Comparison of real versus simulated satellite component architectures. The simulated setup preserves structural and interface fidelity, enabling realistic testing.}
  \label{fig:real_vs_sim}
\end{figure}

\subsection{Satellite Simulation}

We evaluated \attack's feasibility in small satellites using NASA’s NOS3, a comprehensive framework that emulates flight conditions through cFS, simulated hardware drivers, real-time telemetry, and COSMOS ground station tools. This setup mirrors a satellite’s software stack in flight and enables realistic experimentation without physical hardware.

At its core, NOS3 runs on cFS, a modular flight software framework used widely in government, commercial, and academic missions. Applications communicate via a publish–subscribe software bus, where message authentication and access control are not provided, reflecting real-world limitations. In the simulation, each component includes a cFS application (e.g., component\_app.c) that handles commands and telemetry on the software bus. While NOS3 exposes the application source for development, in practice, vendors typically deliver only binaries and interface specifications. As a result, the internal logic of the application—where malicious functionality could be centralized—is opaque to integrators, analogous to component\_app.c being hidden in real hardware. As Figure~\ref{fig:real_vs_sim} illustrates, this mirrors the structure of physical satellites and highlights how supply chain malware can remain invisible during integration.

\subsection{Implementation Details}
This section details the concrete implementation of our malware components within NOS3, focusing on Scenario 5. We describe how the trigger agent and attack agent were constructed to blend into the onboard flight software stack, how they exploited the cFS architecture, and how covert communication and exfiltration channels were implemented. Although the file-based coordination in Scenario 5 is unique, the underlying techniques—cloning vendor-style components, subscribing to legitimate telemetry, and reusing trusted system interfaces—are common to all scenarios.

To ensure compatibility, the trigger agent and attack agent were initially cloned from an example component provided in NOS3\cite{nos3}. This created all the files needed by both components to operate, including the main app, message structures, message IDs, tables, hardware simulations, and ground commands.
Each component adheres to the standard cFS execution lifecycle. Upon launch, they both invoke \texttt{AppInit()} to perform setup tasks, including telemetry packet registration, hardware initialization, and software bus subscriptions.
After initialization, they enter the \texttt{while(CFE\_ES\_RunLoop())} main loop to continuously process software bus messages.

\textbf{Trigger Agent Configuration:} On initialization, the trigger agent subscribes to GNSS telemetry using the \texttt{NOVATEL\_OEM615\_DEVICE\_TLM\_MID} message ID. It then enters a dormant polling loop, examining incoming telemetry until orbital insertion is inferred based on GNSS activity. The trigger agent then writes the string "toggle\_exfil" to the covert FIFO file, triggering the attack agent's exfiltration phase.

Later in the mission, the trigger agent can write "kill" to the FIFO file, which will command the attack agent to cause a crash condition. This is implemented by dereferencing a NULL pointer, causing a segmentation fault that appears to ground control as a random software bug.

\textbf{Attack Agent Configuration:}
On initialization, the attack agent uses \texttt{mkfifo()} to create a file at a predefined path accessible to both the trigger and attack agents. A FIFO was chosen over a regular file because it provides built-in synchronization between processes and automatically erases data once it is read, leaving little or no forensic trace. Communication occurs through simple \texttt{write()} and nonblocking \texttt{read()} calls, which in NOS3 remain invisible to operators since the framework lacks syscall auditing or runtime monitoring. After setup, the attack agent subscribes to satellite housekeeping telemetry, events and services telemetry, software bus statistics telemetry, and time services housekeeping telemetry. These streams provide access to application runtime status, warnings, errors, and system clock data. If a particular subsystem were of interest, the attacker would only need to add the corresponding message ID to begin exfiltrating its telemetry.

Subsequently, the attack agent periodically polls the shared file using a nonblocking \texttt{read()} call. Upon reading "toggle\_exfil", it sets an internal flag and begins sending telemetry messages to the radio for exfiltration. To maintain cover, the attack agent and trigger agent both continue their expected operations.

\textbf{Radio Telemetry Downlink:}
In order to get our exfiltrated telemetry to the ground, we must utilize the satellite's radio transmitter. We abuse the UDP socket maintained by the \texttt{GENERIC\_RADIO} flight application to accomplish this. Normally, this socket is used by cFS to send telemetry from the software bus to the radio hardware. Instead, the attack agent opens the same socket and uses the \texttt{sendto()} system call to inject arbitrary UDP packets containing telemetry. Because UDP sockets in Linux-based systems accept raw buffers and destination addresses, this requires no special privileges beyond access to the socket file descriptor.

Upon receiving any packet on this open port, the radio application immediately forwards it to the \texttt{generic\_radioHardwareModel}, which emulates the physical radio. This forwarding logic performs only basic formatting validation and does not inspect or authenticate packet contents. As a result, any well-formed UDP datagram sent via \texttt{sendto()} is accepted as legitimate telemetry. The hardware model then transmits the data over another UDP socket to the ground station interface.

It is important to note that in NOS3, the radio always transmits to a single default ground station endpoint. Since this cannot be reconfigured at runtime and NOS3 does not simulate directional antennas or frequency hopping, we assume that this default ground station functions as the adversary’s receiver and that the satellite is currently passing overhead. 

These abstractions simplify downlink behavior; however, because our exfiltration reuses the official radio transmission pathway and we are capable of using GNSS to trigger exfiltration, we assume that any packet accepted by the flight radio will be transmitted successfully. 

\subsection{Stealth Characteristics} 
A defining feature of \attack is its ability to conceal malicious activity within normal mission operations. Across scenarios, stealth is enhanced by: (1) using trusted APIs, (2) operating within normal permissions in flight software, and (3) avoiding anomalous telemetry patterns prior to activation. 

These attributes allowed the malware to remain dormant during ground-based testing and to activate only under mission-relevant conditions, enabling long-term persistence without detection in post-launch environments. Although all the attack scenarios employ stealth mechanisms, only \scfive achieves a completely stealth approach (see Table~\ref{tab:attack_results}). We describe the limitations of each scenario in the following:

\scone and \sctwo represent the simplest configurations: a single malicious application with either a static or dynamic telemetry-based trigger.
While these two scenarios can successfully exfiltrate data without being detected once launched, pre-launch detection techniques such as symbolic execution, fuzzing, or control-flow graph analysis could potentially detect the malicious behavior of the components~\cite{arzt2014flowdroid,zhang2014droidsift,wong2016intellidroid,li2015iccta} due to their centralized behavior. 

\scthree and \scfour introduce multiple coordinated components. Separating the trigger and exfiltration functions complicates detection engines relying on centralized behaviors; in other words, single component-based analyses will not work~\cite{BLASCO201888, bosu2017dialdroid,bagheri2015covert,wang2018interapp}. These scenarios can still be identified from their reliance on the software bus for communication in flight, which is visible to ground station operators. Even if communication is obfuscated, if a device transmits at unexpected times, or repurposes commands in ways inconsistent with its documented interface, operators can infer malicious activity. This modularity increased the stealth of \attack but also raised constraints that could ultimately expose it.

\scfive overcomes the limitations of previous scenarios by utilizing multiple components and a communication vector that is invisible to satellite operators, leveraging known limitations such as continuous logging and resource constraints~\cite{schwenk2023board,collins2024merge}. In this sense, the use of a FIFO file for communication enables attackers to execute malicious activities without operators knowing it is happening at all. Accordingly, only scenario five demonstrates stealth; the others do not.
\begin{table*}[ht]
    \centering
    \resizebox{1.8\columnwidth}{!}{
    \renewcommand{\arraystretch}{1.2}
    \begin{tabular}{llllll}
        \hline
        \toprule
            Attack Scenarios & Threat Actor & Trigger & Coordination & Stealth & Countermeasure  \\
            \hline
        \midrule
            \scone &  Solo & Time & - & \ding{55}  & \textbf{Runtime Behavior Monitoring} \\ 
            \sctwo &  Solo & Dynamic & - & \ding{55}  & \textbf{Runtime Behavior Monitoring} \\  
            \scthree  &  Colluding & Time & Software Bus & \ding{55} & \textbf{Software Bus Auth. \& Access Control} \\ 
            \scfour &  Colluding & Dynamic & Software Bus & \ding{55} & \textbf{Software Bus Auth. \& Access Control} \\ 
            \scfive  & Colluding & Dynamic & FIFO File & \ding{51}  & \textbf{Syscall Filtering / OS Hardening}  \\ 
        \bottomrule
        \end{tabular}\vspace{-2mm}
    }
    \caption{Description of the different attack scenarios and their countermeasures. Trigger indicates whether the attack activates based on time or dynamic telemetry. Coordination lists the communication channel between malicious components. Stealth indicates whether the attack avoids detection in both pre-launch analysis and operator-visible channels.}
    \label{tab:attack_results}
\end{table*}

\section{Results} \label{sec:results}

Our simulation demonstrated the successful embedding of stealthy and persistent malware on a small satellite without detection. The following presents key findings across multiple domains, highlighting the implications of supply chain compromise in small satellite missions.

\subsection{Component Coordination}
In all multi-component scenarios, the trigger and attack agents coordinated successfully using either the cFS software bus (\scthree, \scfour) or a covert FIFO file (\scfive). Software bus coordination relied on normal publish–subscribe messaging, while the FIFO channel enabled synchronization through simple file writes and reads, leaving no operator-facing trace.

From the viewpoint of a COSMOS ground station operator, these behaviors were indistinguishable from routine activity. In \scthree and \scfour, coordination messages appeared on the software bus but were disguised with misleading names and descriptions, allowing them to appear consistent with legitimate component activity. In \scfive, covert file-based coordination was invisible to COSMOS, which cannot access the on-board filesystem state. 

Together, these findings confirm that colluding components can reliably coordinate while appearing legitimate, underscoring the difficulty of distinguishing malicious cooperation from normal spacecraft operations.

\subsection{Data Exfiltration}

Across all five \attack scenarios, the attack agent covertly exfiltrated mission telemetry, including housekeeping data, event logs, software bus statistics, and time synchronization to an unauthorized ground station. This was achieved without alerting operators of radio use or leaving incriminating logs, and in real deployments, such an approach could extend to imagery, attitude parameters, or even encryption keys.

In our testing, the COSMOS ground station only recorded what arrived on the configured radio downlink: the timestamp, raw CCSDS packet bytes, XTCE decoding status, and any defined telemetry fields. All packets appeared to originate directly from the radio, with no attribution to the cFS component that generated them. Critically, the satellite had no internal mechanism to log or track when the radio was used for transmission. As a result, if telemetry were exfiltrated to a malicious ground station, operators would see no evidence that it occurred.

\begin{lstlisting}[style=light-terminal, caption={Trigger Software Bus Output (Can Be Disabled)}, label={lst:bus_output}]
EVS Port1 EVS Port1 42/J/TORQUER 14: GENERIC_TORQUER: Device enabled
EVS Port1 EVS Port1 42/J/NAV 13: NOVATEL_OEM615: Enable command received
@[TRIGGER AGENT] GPS Telemetry detected - triggering exfiltration.@
@ATTACK AGENT: Read from FIFO: toggle_exfil@
@ATTACK AGENT: FIFO command received: toggle_exfil@
EVS Port1 EVS Port1 42/J/NAV 14: NOVATEL_OEM615: Device enabled
\end{lstlisting}

\subsection{Deception}

Beyond exfiltration, our simulated malware was able to send arbitrary messages and commands onto the software bus, enabling the generation of fake diagnostic reports, spurious telemetry, or message floods. This activity could mask real anomalies, confuse fault detection, or even trigger unintended behaviors in other applications that subscribe to message IDs that are being flooded. The cFS architecture implicitly trusted all components to publish such messages, giving the malware the same freedom as legitimate applications.

From the COSMOS ground station view, these injected messages blended with legitimate telemetry. This could mislead operators into attributing messages to the wrong component, overwhelm them with filler diagnostics, or even appear as a software bug when commands were sent to other applications—causing crashes or unintended behavior. In each case, situational awareness is degraded, and trust in telemetry eroded. Listing~\ref{lst:arbitrary} shows the attack agent generating fabricated calibration telemetry intended to mislead ground operators.

\begin{lstlisting}[style=light-terminal, caption={Arbitrary logging to confuse GS}, label={lst:arbitrary}]
EVS Port1 EVS Port1 42/1/THRUSTER: Detected deviation from attitude trajectory. Automated thruster correction sequence initiated.@
EVS Port1 EVS Port1 42/1/ATTACK AGENT 28:  [ANOMALY DETECTED] Propellant line pressure drop in thrusters. Recommend disabling thrusters to prevent cascade failure.
\end{lstlisting}

\subsection{Persistence and Denial of Service}

In each scenario, \attack succeeded in escalating from data exfiltration to denial of service (DoS), triggered by either static or sensor-based commands. Upon activation, the attack agent deliberately crashed via a null pointer dereference, producing segmentation faults that mimic common software failures. These crashes can be looped or repeated to sustain disruption and complicate recovery. To operators, such events appear as plausible application bugs, providing the attacker with both plausible deniability and the ability to suppress forensic evidence—particularly dangerous when the satellite is out of contact of a ground station for analysis.

Alternatively, \attack remains persistent, continues to exfiltrate telemetry indefinitely while blending with normal operations. In this mode, the attack agent processed bus messages and generated housekeeping data, maintaining the appearance of legitimacy and leaving operators with no anomalies to investigate. In addition, the trigger agent could further reissue commands based on orbital conditions, enabling adaptive control over when exfiltration pauses or resumes. Together, these capabilities allowed \attack to conceal activity, disrupt response, and sustain long-term presence on the satellite.

\begin{lstlisting}[style=light-terminal, caption={NOS3 Flight Software Boot Banner}, label={lst:reset}]
Executing Denial of Service...
@cFS Restarting...@
...
@CFE_PSP: Reset Type: PO@
CFE_PSP: Default Reset SubType = 1
CFE_PSP: Default CPU ID = 1
CFE_PSP: Default Spacecraft ID = 42
OS_Posix_GetSchedulerParams():191: Policy 1: available, min-max: 1-99
\end{lstlisting}

\subsection{Identifying Novel Strategies}

Developing \attack involved utilizing multiple tactics, techniques, and procedures (TTP). After defining our attack pipeline, we mapped each TTP used over each stage of the attack with the standardized SPARTA matrix~\cite{aerospace2022sparta}. Table~\ref{tab:sparta} presents the complete list of TTPs that we utilized, showing that our attack methodology aligns with realistic attack scenarios. Although the list is comprehensive and accurately represents our design decisions, it overlooks a vital technique: the colluding behavior of compromised components. This novel technique is key to our findings because it allows \attack to bypass testing and detection. By distributing malicious logic, \attack avoids centralized signatures, increases persistence, and achieves a higher level of stealth. We successfully utilized this technique in \scthree, \scfour, and \scfive, demonstrating its effectiveness in sustaining covert activity during mission operations. We informed the SPARTA team of this colluding-component technique, and they have incorporated it into the matrix as \href{https://sparta.aerospace.org/technique/DE-0012/}{Component Collusion} under Defense Evasion with ID DE-0012.

\begin{table}[t]
    \centering
    \scalebox{0.7}{
    \begin{tabular}{ll>{\raggedright\arraybackslash}p{0.75\linewidth}}
        \toprule
        \textbf{Category} & \textbf{TTP} & \textbf{Name} \\
        \midrule
        \hline
        \multirow{2}{*}{\textbf{Recon.}} 
        & REC-0001.01 & Gather Spacecraft Design Information: Software Design \\
        & REC-0001.04 & Gather Spacecraft Design Information: Data Bus \\
        \midrule
        \multirow{3}{*}{\textbf{I. Access}} 
        & IA-0010 & Unauthorized Access During Safe-Mode \\
        & IA-0001.02  & Compromise Supply Chain: Software Supply Chain \\ 
        & IA-0009.0 & Trusted Relationship: Vendor \\
        \midrule
        \multirow{5}{*}{\textbf{Execution}} 
        & EX-0002 & Position, Navigation, and Timing Geofencing \\ 
        & EX-0010 & Malicious Code \\ 
        & EX-0009.01 & Exploit Code Flaws: Flight Software \\
        & EX-0008 & Time Synchronized Execution \\ 
        & EX-0013 & Flooding \\
        \midrule
        \textbf{Persistence} & PER-0002.01 & Backdoor: Hardware Backdoor \\
        \midrule
        \textbf{Exfiltration} & EXF-0003.02 & Signal Interception: Downlink Exfiltration \\
        \midrule
        \multirow{4}{*}{\textbf{Impact}}
        & IMP-0001  & Deception \\
        & IMP-0002  & Disruption \\
        & IMP-0003  & Denial \\
        & IMP-0006  & Theft \\
        \midrule
        \rowcolor{red!10}
        \textbf{Defense Evasion} & DE-0012 & Component Collusion \\ 
        \bottomrule
    \end{tabular}
    }
    \caption{SPARTA techniques involved in \attack. Each technique corresponds to a specific tactic, technique, or procedure (TTP) used in the adversarial model. Recon. stands for Reconnaissance. I. Access stands for Initial Access. The last row is our new TTP, Component Collusion.}
    \label{tab:sparta}
\end{table}

\section{Countermeasures} \label{sec:counter}

\vspace{-1mm}

To date, there are no public operational IDS/IPS solutions deployed on satellites. While an experimental NOS3–based IDS has been proposed~\cite{lolive2025embeddedIDS}, its probe–based design is narrowly scoped and would not detect the stealth techniques demonstrated by \attack. This gap highlights the need for lightweight, embedded defenses tailored to satellites.

In the following, we outline the weaknesses exploited by \attack and recommend countermeasures to strengthen resilience. Table~\ref{tab:attack_results} summarizes the attack scenarios and the most effective mitigations. We present practical defenses for NOS3 and cFS, followed by broader recommendations for hardening satellite systems against supply chain threats.

\subsection{NOS3 \& cFS Specific Recommendations}
These strategies target structural weaknesses specific to NOS3 and the Core Flight Software ecosystem.

\vspace{-2mm}
\subsubsection{Improve Runtime Behavior Monitoring}
As shown in Table~\ref{tab:attack_results}, \scone and \sctwo demonstrate malware that appears to operate as expected. The components use normal APIs and legitimate Software Bus subscriptions, but their behavioral patterns—such as frequency, volume, timing, or system calls—are abnormal. Existing IDS prototypes~\cite{lolive2025embeddedIDS} are limited to probing command traffic and generic bus features, and therefore miss covert techniques like hidden FIFO channels or excessive internal telemetry subscriptions that never alter ground-command flows.

\textbf{Recommendation:} We propose lightweight runtime monitoring or behavioral auditing agents within the flight software. These could track system call usage per application, monitor file and network access, and observe message bus activity for anomalies in frequency or subscriptions. Even simple heuristics, such as anomalous spikes in traffic or radio transmission, could serve as effective compromise indicators.

\textbf{Code Implementation:}
To illustrate feasibility, we implemented a prototype runtime monitor in the NOS3 simulation. Our proof-of-concept added a lightweight message-rate limiter inside the attack agent: internal counters tracked message transmissions within short windows, and once a threshold (e.g., 100 messages) was exceeded, the component entered a throttled state, suppressing further transmissions. A warning event was logged whenever throttling was triggered, providing visibility and accountability for runtime violations. Although tested in simulation, this mechanism models a containment response that could be ported to real cFS components and extended to audit radio usage, command activity, or other behaviors indicative of covert malware.

\subsubsection{Software Bus Authentication and Access Control} 
As seen in Table~\ref{tab:attack_results}, \scthree and \scfour exploit a lack of authentication and access control on the Software Bus to coordinate their actions and exfiltrate data. The trigger and attack agents can subscribe to system-wide telemetry streams, including mission-critical data, without verification, and can publish arbitrary messages and commands at will. This lack of enforcement enables malicious components to monitor sensitive information and issue covert signals while appearing legitimate within the flight software ecosystem.

\textbf{Recommendation:} Introduce Software Bus access control lists and message authentication. Each module should be required to register its topics and actions in a manifest, verified at runtime. This change would prevent the attack module from simply subscribing to private telemetry streams.

\textbf{Code Implementation:} To demonstrate feasibility, we modified a cFS application to enforce an allowlist of authorized message IDs. Before completing any Software Bus subscription request, the application checks the requested ID against its predefined policy. Unauthorized requests are denied and logged, preventing components from silently subscribing to restricted telemetry streams. Although this prototype implements policy enforcement locally within one component, it illustrates how access control could be integrated into the cFS framework itself. Future extensions could include centralized manifests, per-component configuration files, or cryptographic signatures to provide stronger system-wide enforcement.

\vspace{-1mm}
\subsubsection{Restrict Unmonitored System Calls} 
\scfive, which uses covert FIFO–based signaling, depends heavily on access to system calls such as \texttt{mkfifo()}. As shown in Table~\ref{tab:attack_results}, restricting or filtering access to such calls is the most effective countermeasure, blocking covert coordination entirely.

\textbf{Recommendation:} System call usage should be restricted at runtime to reduce the stealth and effectiveness of malicious components. This can be implemented using OS-level security mechanisms—such as \texttt{seccomp} in Linux or equivalent sandboxing frameworks in real-time operating systems—configured to enforce per-component allowlists of permitted system calls. Under such a policy, third-party modules attempting to invoke sensitive functions (e.g., \texttt{mkfifo()} or \texttt{sendto()}) would be blocked before covert channels or unauthorized exfiltration could occur. More broadly, satellite operating systems should be hardened to minimize unnecessary kernel interfaces exposed to flight software, shrinking the attack surface available to supply-chain malware.

\textbf{Code Implementation Attempt:} While full seccomp integration into NOS3 proved challenging due to compatibility and initialization issues, we explored a conceptual implementation. A syscall filter would be invoked during component initialization, defining a minimal set of allowed system calls and denying all others by default. Any disallowed syscall—such as those used to open sockets or files—would trigger immediate termination of the process. Although tested only in simulation, this mechanism illustrates how lightweight, per-component syscall restrictions could be adopted in cFS-based systems or hardened OS deployments to improve software isolation and reduce attack surface in small satellites.

\subsection{General Security Recommendations}

\attack exploited trusted software interfaces, diagnostic roles, system calls, and insufficient monitoring. These vulnerabilities are not unique to cFS, but reflect broad weaknesses in small–satellite development practices, amplified by the lack of deployable detection tools.

\vspace{-1mm}
\subsubsection{Zero Trust Flight Software Design} 
In our scenarios, malware succeeded because components were implicitly trusted with broad access to telemetry, files, and network functions. This model makes it trivial for malicious code to blend in and misuse legitimate interfaces for covert exfiltration. 

\textbf{Recommendation:} Flight software must be built on the assumption that any module could be compromised, with defenses layered into its design. Developers can enforce least privilege by declaring which telemetry streams, file paths, and network interfaces each application may access, denying everything else by default. Manifest-driven access control ensures modules only subscribe or publish to approved topics. In addition, tracking message rates, command frequency, or socket use could provide early warning of abnormal behavior. Together, these practices create stronger boundaries and make covert misuse harder to hide, even in the absence of a space-ready IDS. It's essential that developers build security into flight software architecturally rather than apply fixes to architecturally weak software.

\vspace{-1mm}
\subsubsection{Supply Chain Verification and Transparency} 
The adoption of COTS creates blind spots when third-party parts are integrated with little visibility into internal logic. Once on-orbit, these modules are effectively immutable—difficult to patch, monitor, or replace—making any hidden payloads a long-term liability that \attack took advantage of.

\textbf{Recommendation:} Flight software teams should require verifiable provenance and reproducibility for all integrated components. This includes vendor disclosure of sourcing, reproducible build pipelines, and post-installation analysis of delivered binaries. Maintaining a software bill of materials and linking it to integration testing would provide traceability across the supply chain and support forensic analysis after launch. These measures would help ensure that black-box components do not become silent points of failure.

\subsubsection{Operational Baselines and Training} 
Even with technical safeguards, detection ultimately depends on operators recognizing when a system is behaving abnormally. Yet ground crews are rarely trained to distinguish benign anomalies from the subtle footprints of coordinated malware. Without reference baselines, covert data exfiltration or unusual diagnostic activity may pass unnoticed.

\textbf{Recommendation:} Develop operator playbooks that document baseline telemetry patterns, expected message rates, and normal process behavior. Train teams to flag deviations such as unexpected messaging in the software bus. Regular tabletop exercises simulating supply-chain compromise can prepare operators to respond decisively under uncertainty. Embedding this operational awareness into mission procedures ensures that security is not solely a design concern but a continuous in-flight practice.

By treating security as a baseline design requirement rather than an afterthought, missions can raise the cost of compromise and reduce the likelihood that covert supply-chain malware will persist undetected in orbit.
\section{Discussion}\label{sec:discuss}

Our study shows that supply chain attacks can operate undetected in small satellite environments. Across five scenarios, we demonstrated that malicious components, whether alone or in collusion, can coordinate, exfiltrate telemetry, and abuse implicit trust in the flight software to avoid detection and compromise mission success. These findings validate our threat model: trusted third-party components integrated into modular satellite architectures can sustain stealthy, persistent malware operations that bypass both pre-launch validation and in-orbit monitoring.

Unlike prior satellite security studies that focused on isolated exploits or assumed insider access to core software, our work demonstrates a realistic multi-component supply chain attack within NASA’s NOS3 environment. Our findings underscore critical implications for stakeholders: vendors must strengthen assurance practices for plug-and-play modules, developers of flight software need to adopt runtime monitoring and access controls, operators cannot assume that well-formed telemetry is truthful, and end users face mission-critical risks when data integrity is silently undermined. At the same time, it is important to acknowledge the limitations of our simulation in order to clarify the scope of these findings and outline directions for future research.

\textbf{Limitations:}
Our simulation demonstrates the feasibility of embedding covert malware within a small satellite system using NASA’s NOS3, a widely used framework in NASA, industry, and academia. However, NOS3 cannot fully capture the operational dynamics and physical constraints of orbital missions. Effects such as radiation-induced faults, thermal stress, and variable link quality remain outside its scope.

Our exfiltration method exploited telemetry injection through the legitimate radio link, successfully reaching the COSMOS ground station without new connections or logs. While this validates the covert channel in simulation, real-world communication involves orbital dynamics, scheduled passes, and fluctuating link quality, which could complicate or delay data exfiltration. Further study is needed to assess persistence under operational conditions

\textbf{Need for Community:}
Our findings reveal an urgent need for deeper engagement between stakeholders, including satellite developers and security researchers. Despite increasing reliance on small satellites in defense, climate science, and commercial applications, their software development lifecycle often lacks rigorous security validation. Addressing this problem will require standardized frameworks for testing, shared benchmarks, and modular detection tools that can be adapted across missions. Open-source environments like NOS3 provide a strong foundation, but community efforts must evolve toward complete testbeds with anomaly detection, telemetry validation, and attack simulation capabilities.

Equally important is broader collaboration between academic, industry, and government stakeholders to develop guidelines, tools, and policies. A public repository of attack patterns, malware examples, and secure software templates could accelerate the adoption of secure practices, while integrating formal verification, compliance checks, and continuous threat modeling into the lifecycle would strengthen resilience. Together, these measures would help ensure that future satellites are better equipped to withstand covert, supply-chain–borne threats without compromising the performance or modularity that make small satellites so valuable.

\section{Conclusion} \label{sec:conclude}

Small satellites are increasingly complex, autonomous platforms that depend on third-party components, amplifying the risk of covert software compromise. \attack, implemented in NASA’s NOS3 environment, provides the first end-to-end demonstration of coordinated, multi-component malware in this setting, showing how trusted components can collude to exfiltrate telemetry, persist across reboots, and disrupt operations using legitimate interfaces.

By introducing a taxonomy of escalating stealth and complexity, validating sensor-aware and colluding malware behaviors, and distilling these into actionable hardening strategies, our work underscores both the feasibility of supply-chain compromises in space systems and the urgent need for resilience.

Embedding zero-trust principles, authenticated telemetry, anomaly detection, and post-launch validation throughout the satellite life cycle and extending platforms like NOS3 into cybersecurity testbeds will be essential in securing the next generation of small-satellite missions.

\section*{Ethical Considerations}
\noindent\textbf{Stakeholders.}  
This work potentially affects a wide range of stakeholders. \textit{Satellite developers and operators} may be impacted, as our simulated scenarios highlight risks that could plausibly apply to their systems. \textit{Vendors of third-party components} are relevant stakeholders, since our model assumes the possibility of insider access at this level. \textit{NASA and NOS3 maintainers} are directly connected, given that our work was performed entirely in the NOS3 environment and included engagement with the development team. \textit{Security researchers} stand to benefit from the knowledge, methodologies, and proposed mitigations provided by this study. At the same time, \textit{adversaries} are a stakeholder of concern, as there is a risk that publication might lower barriers to attack if misused. Finally, \textit{society at large} is an indirect stakeholder, since satellites underpin critical functions such as communications, Earth observation, and national security.

\noindent\textbf{Ethical Principles.}  
We evaluated our research through the lens of the Menlo Report. \textit{Beneficence:} Our study reveals realistic supply-chain risks and provides concrete mitigations to help defend against them, thereby contributing net benefit to the community. \textit{Respect for Persons:} No human subjects were involved, and we respected community norms by avoiding live satellites, vendor-specific systems, or production infrastructure. All experiments were confined to the NOS3 simulation environment. \textit{Justice:} The benefits of this work apply broadly to the satellite ecosystem and do not unfairly burden any single group. \textit{Respect for Law and Public Interest:} The research was conducted lawfully, with no terms of service or operational systems violated. Engagement with NASA ensures that the outcomes align with public-interest goals rather than undermine them.

\noindent\textbf{Potential Harms.}  
We identified several possible harms. \textit{Tangible harms} were avoided since all work was performed in simulation, with no disruption to operational satellites or use of mission data. However, there is a risk of \textit{knowledge misuse}, as adversaries could attempt to adapt concepts described in this paper. A further concern is \textit{reputation risk} to vendors, who may feel implicated despite our not naming or testing any specific COTS products. Lastly, \textit{researcher wellbeing} was considered, though our work involved no disturbing content or hazardous environments.

\noindent\textbf{Mitigations.}  
Several steps were taken to reduce risk. All work was restricted to the NOS3 simulation environment. We deliberately avoided identifying specific vendors or analyzing real spacecraft systems. Mitigation strategies were shared and discussed with the NASA NOS3 team to ensure that our contributions advance defensive capabilities. In publication, we balance academic rigor with caution: the technical details are sufficient for reproducibility and scholarly review, but do not include step-by-step exploit code that would enable immediate weaponization.

\noindent\textbf{Decision to Proceed and Publish.}  
The decision to proceed stemmed from a clear blind spot in prior work: the lack of realistic, implemented demonstrations of supply-chain malware in satellites. Exploring this issue in a safe environment (NOS3) allows the community to better understand these risks. The decision to publish was reached because the work advances collective defense by (1) exposing how trusted but opaque code can be abused, (2) demonstrating systemic constraints that limit defenses, and (3) offering practical countermeasures for satellite teams. On balance, the ethical benefits — raising awareness, strengthening resilience, and providing mitigations — outweigh the potential harms, particularly given the exclusively simulated nature of our experiments.

\noindent\textbf{Concluding Note.}  
We recognize the principle that sometimes the most ethical decision is not to conduct or publish research. In this case, we judged that proceeding was ethically appropriate because all experiments were confined to a simulator, no live assets were affected, and the primary contribution of this work is to improve resilience in a domain where practical, system-level demonstrations of supply-chain malware have been scarce.

\bibliographystyle{unsrt}
\bibliography{strings,bib}

\end{document}